# Coupling multi-space topologies in 2D ferromagnetic lattice

Zhonglin He, Wenhui Du, Kaiying Dou, Ying Dai*, Baibiao Huang, Yandong Ma*

School of Physics, State Key Laboratory of Crystal Materials, Shandong University, Shandanan Street 27, Jinan 250100, China

*Corresponding author: daiy60@sina.com (Y.D.); yandong.ma@sdu.edu.cn (Y.M.)

**Abstract**

Topology can manifest topological magnetism (e.g., skyrmion and bimeron) in real space and quantum anomalous Hall (QAH) state in momentum space, which have changed the modern conceptions of matter phase. While the topologies in different spaces are widely studied separately, their coexistence and coupling in single phase is seldomly explored. Here, we report a novel phenomenon that arises from the interaction of topological magnetism and band topology, the multi-space topology, in 2D ferromagnetic lattice. Based on continuum theory and tight-binding model, we reveal that the interconnection between skyrmion/bimeron and QAH state generates distinctive localized chiral bound states (CBSs). With moderating topological magnetism through magnetic field, the multi-space topologies accompanied with different CBSs can be reversed, facilitating the coupling of multi-space topologies. By performing first-principles and atomic spin model simulations, we further demonstrate such multi-space topologies and their coupling in monolayer $Cr_2NSb$. These results represent an important step towards the development of multi-space topological phenomena in 2D lattice.

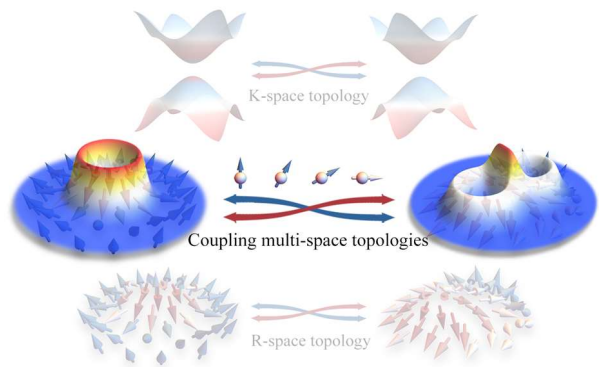

TOC

**Introduction**

Topology, described by geometric phase, is a central theme of current condensed-matter research [1-4]. Physically, topology can exist in both real (R) and momentum (K) spaces [5,6]. Manifestation of topology in K-space generates band topology, e.g., quantum anomalous Hall (QAH) state [7]. It is characterized by nonzero quantized K-space geometric phase, i.e., Chern number $\mathbb{C}$ [8, 9], which guarantees the stability of chiral edge state (CES) supporting dissipationless current flows, ushering in dissipation-free memory and logic devices [10,11]. Topology manifested in R-space gives rise to topological magnetism, such as skyrmion and bimeron [12-14]. These topologically protected spin textures harbor nonzero quantized R-space geometric phase, i.e., topological charge $\mathbb{Q}$ [15]. Such R-space topology protects skyrmion and bimeron from annihilation and enables their quasi-particle behaviors when interacting with external probes, rendering them great potential as information carriers for high-performance spintronic devices [16-18].

While both band topology and topological magnetism have undergone rapid development and attracted broad attention, an intriguing but rarely explored scenario is K-space + R-space topology (that is, multi-space topology with coupling K-space topology to R-space topology) [19-23]. In principle, K-space and R-space topologies in condensed-matter physics do not mutually exclusive each other. Instead, they may coexist or even couple with each other, with two-dimensional (2D) magnetic material as a bridge [24-26]. Obviously, the K-space + R-space topology scenario is of significant interest at both fundamental and applied levels, as it can change the topological properties of the host materials and bridge the physics of K-space and R-space topologies. Despite the fundamental interest, multi-space topology is rarely investigated owning to the scarcity of the coexistence of K-space and R-space topologies. Actually, up to now, the coupling of multi-space topologies has not been proposed.

In this work, we report a new phenomenon that arises from the interaction of topological magnetism and band topology, the multi-space topology, in 2D ferromagnetic lattice. Based on continuum theory and tight-binding (TB) model, we reveal that the interconnection between skyrmion/bimeron and QAH state generates distinctive chiral bound states (CBSs). Benefiting from tuning topological magnetism through magnetic field, the multi-space topologies associated with different CBSs can be switched, thereby generating the coupling of multi-space topologies. Importantly, this process is accompanied with the space-shift of the CBSs. Using first-principles calculations and atomic spin model simulations, these multi-space topologies and their strong coupling are further validated in monolayer $Cr_2NSb$. The underlying physics is discussed in detail. These findings not only enrich topological physics, but also offer a pathway towards the control of multi-space topology.

**Results and Discussion**

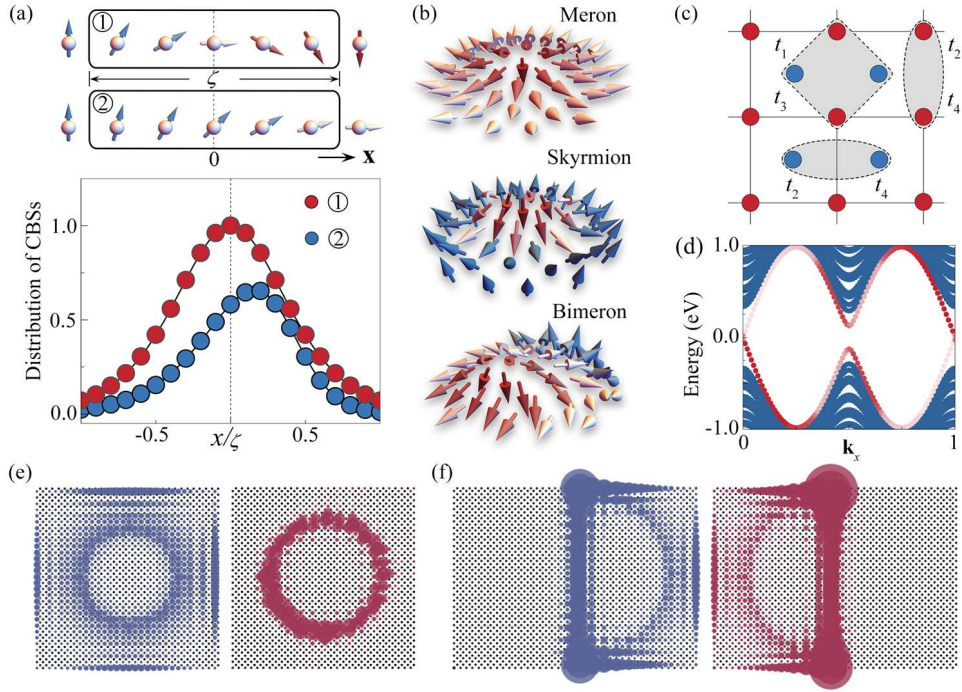

**Fig. 1.** (a) Top: Two typical boundaries of topological magnetic domains. Bottom: Distributions of CBSs for the two typical boundaries as a function of $x$. (b) Schematic illustrations of spin distributions for meron, skyrmion and bimeron. (c) Sketch of the TB model for the 2D square lattice. (d) Edge spectra of the OP phase. CES within the bulk gap is depicted by the red circles, where the intensity of color is proportional to the density of states. (e) Distributions of CBS-M (blue) and CBS-SK (red). (f) Distribution of CBS-BM, wherein red and blue are utilized to distinguish the CBSs localized around the meron and antimeron in bimeron, respectively.

K-space topological domain of QAH state is characterized by Chern number $\mathbb{C}$. Considering that $\mathbb{C}$ usually depends on spin orientations [$\mathbb{C}(\mathbf{S}_z) = \begin{cases} 1, \mathbf{S}_z > 0 \\ 0, \mathbf{S}_z = 0 \\ -1, \mathbf{S}_z < 0 \end{cases}$, where $\mathbf{S}_z$ describes the distribution of out-of-plane (OP) spin component], K-space topological domains are associated with R-space magnetic domains. The CESs would emerge near the boundaries of R-space magnetic domains due to the bulk-boundary correspondence [22], which are manifested as CBSs in R-space. Quietly naturally, it is wondering whether

CBSs can be localized near the boundaries of topological magnetic domains (e.g., skyrmion and bimeron) and thus utilized to describe the interactions between topological magnetism and QAH state. To investigate the characteristics of CBSs near the boundaries of topological magnetic domains, we employ the Hamiltonian [27, 28]

$$H = v_F(\mathbf{k}_x \sigma^y - \mathbf{k}_y \sigma^x) + H_Z(\mathbf{S}_z). \tag{1}$$

Here, the first term is Rashba Hamiltonian characterizing massless Dirac electrons, where $\mathbf{k}_i$ ($i = x, y$) is momentum operator, $v_F > 0$ is the Fermi velocity and $\sigma^i$ ($i = x, y, z$) denotes Pauli matrices for spin. The second term opens a band gap via Zeeman field and the form is shown as **Eq. (S3)**. $\mathbf{S}_z = \pm 1$ relates to OP ferromagnetic (FM) phase (OP-↑ and OP-↓ phases), while $\mathbf{S}_z = 0$ represents in-plane (IP) FM phase (IP phase). According to **Eqs. (1) and (S4)**, OP-↑ (↓) phase corresponds to QAH state with $\mathbb{C} = +1$ (-1), while IP phase exhibits a trivial gap. For convenience, unless otherwise stated, IP phase refers to FM phase with spins oriented along **x** axis in the following.

When the boundary scale is significantly small as compared with topological magnetic domains, the spin textures of topological magnetism vary slowly along the direction perpendicular to the boundary. The direction along the boundary is set to **x**, then $\mathbf{k}_y$ can be considered as a good quantum number. The Hamiltonian of **Eq. (1)** can be simplified in one dimension with $\mathbf{k}_y = 0$ [27, 29] under the continuum approximation

$$H = \begin{pmatrix} J\mathbf{S}_z(x) & -v_F \partial_x - \frac{\Delta}{2}\mathbf{S}_x(x) \\ v_F \partial_x - \frac{\Delta}{2}\mathbf{S}_x(x) & -J\mathbf{S}_z(x) \end{pmatrix}. \tag{2}$$

The CBSs can be obtained by solving $H\Psi = 0$. For simplicity, both $J$ and $v_F$ are set to 1 and $\Delta$ is set to 0.2. The topological magnetic domains can be roughly classified as IP, OP-↑ and OP-↓ domains. This results in two typical boundaries between these domains, as depicted in the top panel of **Fig. 1(a)**. $\zeta$ represents the width of the boundaries. For ①-boundary, the OP spin distribution is expressed by $\mathbf{S}_z(x) = \tanh\left(-\frac{x}{\zeta}\right)$, which describes a continuous rotation of $\mathbf{S}_z(x)$ from +1 to -1. The distribution of CBS follows the relation of $|\Psi(x)|^2 \propto \mathrm{sech}(\frac{x}{\zeta})^{2\zeta}$, and as shown in the low panel of **Fig. 1(a)**, it is strongly localized around the center of boundary. While for ②-boundary, a continuous rotation of $\mathbf{S}_z(x)$ from +1 to 0 yields $\mathbf{S}_z(x) = \frac{\tanh(-\frac{x}{\zeta})+1}{2}$. The corresponding CBS distributes following $|\Psi(x)|^2 \propto \frac{e^{2x/\zeta}}{(1+e^{2x/\zeta})(4+\Delta^2 e^{2x/\zeta}(2+e^{2x/\zeta}))\zeta^2}$, which is also strongly localized around the center of boundary [low panel of **Fig. 1(a)**], but significantly different from that of ①-boundary. Accordingly, CBSs can describe the interactions of topological magnetism and QAH state, i.e., characterizing the multi-space topology.

To gain deeper insight into the CBSs aroused from multi-space topology, we further construct a TB model of a 2D FM square lattice [see **Fig. 1(c)**], which is usually utilized to explore the K-space topology [30, 31]. The Hamiltonian of this lattice can be formulated as

$$H(\mathbf{k}) = 2t_1 \cos\frac{\mathbf{k}_x}{2}\cos\frac{\mathbf{k}_y}{2}\tau^x + t_2(\cos\mathbf{k}_x + \cos\mathbf{k}_y) + t_3\tau^y(\sigma^y \cos\frac{\mathbf{k}_y}{2}\sin\frac{\mathbf{k}_x}{2} - \sigma^x \cos\frac{\mathbf{k}_x}{2}\sin\frac{\mathbf{k}_y}{2}) \\ + t_4\tau^z(\sigma^y \sin\mathbf{k}_x - \sigma^x \sin\mathbf{k}_y) + t_5\sigma^z \quad (3)$$

Here, $\tau^i$ ($i = x, y, z$) denote the Pauli matrices of sublattice degree of freedom. $t_1$, $t_2$, $t_3$, $t_4$ and $t_5$ are coefficients of nearest-neighbor (NN) hopping, next nearest-neighbor (NNN) hopping, intrinsic SOC, Rashba SOC and FM exchange coupling, respectively, as illustrated in **Fig. 1(c)**. According to **Eq. (S4)**, the OP phase is a Chern insulator with $|\mathbb{C}|=1$ sustaining CESs within the bulk gap [see **Figs. 1(d)** and **S1**], while the IP phase is a trivial insulator characterized by $\mathbb{C}=0$ (see **Supplementary Note 3**).

Upon introducing topological magnetic domain, an addition term describing the interaction of electrons and spin textures should be included in **Eq. (3)**

$$H_0(\mathbf{k}) = \lambda_i \sigma^i \mathbf{S}_i \quad (4)$$

with $\lambda_i$ ($i = x, y, z$) (see **Supplementary Note 3**). Here, three typical spin textures are considered, i.e., skyrmion, bimeron and meron. As shown in **Fig. 1(b)**, skyrmion presents ①-boundary, meron shows ②-boundary, and bimeron presents both boundaries. It is convenient to adopt some idealized approximations [29, 32] for spin distributions of the two boundaries (see **Supplementary Note 3**).

By solving the TB model on the basis of **Eq. (S7)**, the distributions of CBSs for these three spin textures are obtained, which are shown in **Figs. 1(e, f)**. From **Fig. 1(e)**, it can be seen that the CBSs of meron (i.e., CBS-M) and skyrmion (i.e., CBS-SK) are both predominantly localized near the boundaries. While for bimeron, as shown in **Fig. 1(f)**, its CBS (CBS-BM) is localized near both ①- and ②-boundaries. And it's apparent that the characteristics of the corresponding CBSs for these three spin textures are significantly different. These facts confirm that the localized CBSs indeed can be aroused from the interaction of topological magnetism and band topology, and thereby can describe the multi-space topology.

It is interesting to note that skyrmion and bimeron are stablised in easy-axis and easy-plane 2D magnets, respectively. By manipulating spin orientation of 2D magnets through magnetic field, the skyrmion and bimeron can be switched. Such skyrmion-bimeron switching has been well established in previous works [33, 34]. Benefiting from tuning skyrmion-bimeron switching through magnetic field, the multi-space topologies associated with different CBSs can be switched. This would generate the coupling of multi-space topologies. And as shown in **Figs. 1(e, f)**, this process is accompanied with the space-shift of the CBSs.

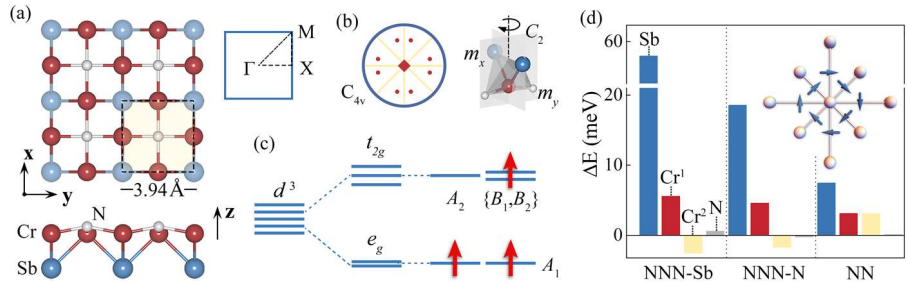

**Fig. 2.** (a) Top and side views of the crystal structure of monolayer $Cr_2NSb$. Insect in (a) is the 2D Brillouin zone. (b) Equatorial plane projection of the point group for monolayer $Cr_2NSb$ and the coordinate environment of Cr atoms. (c) Splitting of $d$ orbitals for Cr atoms. (d) Atomic-layer-resolved localization of the DMI-associated SOC energy $\Delta E$ for monolayer $Cr_2NSb$. Insect in (d) presents the DMI vectors (blue arrows) between the NN and NNN Cr atoms. $Cr^1$ and $Cr^2$ denote the two Cr atoms in the unit cell.

Having demonstrated the feasibility of multi-space topologies and their coupling in 2D FM lattice, we next discuss its realization in a real material of monolayer $Cr_2NSb$. **Fig. 2(a)** displays the crystal structure of monolayer $Cr_2NSb$, which exhibits a square lattice with the space group P4mm (No. 99). It consists of three atomic planes stacked in the sequence of N-Cr-Sb. The lattice parameter is optimized to be 3.94 Å, agreeing well with the previous work [35]. The stability of monolayer Cr$_2$NSb is confirmed by phonon calculations and ab initio molecular dynamics simulations (**Fig. S3**).

The electronic configuration of isolated Cr atom is $3d^5 4s^1$. By donating three valance electrons to the surrounded N and Sb atoms, the oxidation state of Cr ion is +3, presenting an electronic configuration of $3d^3 4s^0$. Under a tetrahedral crystal field, the $d$ orbitals of Cr split into two groups: the higher triplet $t_{2g}$ orbitals and the lower doublet $e_g$ orbitals. The distortion of tetrahedral geometry under the local point group of C$_{2v}$ further splits $e_g$ and $t_{2g}$ orbitals, as illustrated in **Fig. 2(c)**. As a result, the Cr atom exhibits $e_g^2 t_{2g}^1$ configuration, which would give rise to the magnetic moment of 3 $\mu_B$. As expected, our calculations reveal that monolayer Cr$_2$NSb favors a spin-polarized state, with a magnetic moment of 6 $\mu_B$ per unit cell. The magnetic moment is mainly localized on the two Cr atoms. To get further insight into the magnetic properties of monolayer Cr$_2$NSb, we employ a 2D atomic spin Hamiltonian, which can be written as

$$H = -J_{NN} \sum_{\langle i,j \rangle} (\mathbf{S}_i \cdot \mathbf{S}_j) - J_{NNN}^X \sum_{X, \langle\langle i,j \rangle\rangle} (\mathbf{S}_i \cdot \mathbf{S}_j) - K \sum_{\langle i \rangle} (\mathbf{S}_i^z)^2 \\ - \sum_{\langle i \rangle} \mathbf{B} \cdot \mathbf{S}_i - \sum_{\langle i,j \rangle} \mathbf{D}_{NN,ij} \cdot (\mathbf{S}_i \times \mathbf{S}_j) - \sum_{\langle\langle i,j \rangle\rangle} \mathbf{D}_{NNN,ij}^X \cdot (\mathbf{S}_i \times \mathbf{S}_j). \quad (5)$$

Here, spin vector $\mathbf{S}_i$ represents the local magnetic moment at $i^{\text{th}}$ Cr site, with $\mathbf{S}_i^z$ donating the OP component. The summations $\langle i \rangle$, $\langle i,j \rangle$ and $\langle\langle i,j \rangle\rangle$ run over all Cr atoms, NN and NNN Cr atomic pairs, respectively. $X$ = N (Sb) specifies the NNN Cr atomic pairs connected by N (Sb) atoms. $J_{NN}$, $J_{NNN}$, $K$, $\mathbf{B}$, $\mathbf{D}_{NN,ij}$ and $\mathbf{D}_{NNN,ij}$ describe the NN isotropic exchange interaction, NNN isotropic exchange interaction, single-ion anisotropy, external magnetic field, NN and NNN Dzyaloshinskii-Moriya interaction (DMI), respectively. The calculated magnetic parameters are listed in **Table S2**. The positive $J$ suggests FM coupling for monolayer Cr$_2$NSb, which is enhanced by Hund's rule interaction due to the empty $t_{2g}$ orbitals [36]. The calculated magnetic anisotropy $K$ = -1.36 meV implies that monolayer Cr$_2$NSb prefers IP spin orientation.

For stabilizing topological magnetism, DMI usually plays a vital role. DMI is antisymmetric and relies on vector $\mathbf{D}_{ij}$. As shown in **Fig. 2(a)**, the mirror planes halve the NN Cr-Cr pair and include the NNN Cr-$X$-Cr pair, which constrain $\mathbf{D}_{NN,ij} = d_{NN}(\mathbf{u}_{ij} \times \mathbf{z}) + d_z \mathbf{z}$ and $\mathbf{D}_{NNN,ij}^X = d_{NNN}^X (\mathbf{u}_{ij} \times \mathbf{z})$ [37]. $\mathbf{u}_{ij}$ and $\mathbf{z}$ are the unit vectors from sites $i$ to $j$ and along $\mathbf{z}$ direction, respectively. Based on the previous works [26, 38], $d_z$ plays a negligible influence and thereby is neglected. The positive $d_{NN}$ and $d_{NNN}^X$ indicate the clockwise distributions of DMI vectors, as illustrated in **Fig. 2(d)**. To gain more insight into the DMI, we calculate the atomic-resolved localization of DMI-associated SOC energy ΔE. As shown in **Fig. 2(d)**, for the case of NN DMI, both adjacent Sb atom and magnetic Cr atom act as SOC-active sites to induce spin-orbit scattering necessary for DMI, corresponding to the combination of Rashba and Fert-Levy mechanisms [39, 40]. For NNN DMI, it is mainly contributed by the Sb atomic layers, which is related to the Fert-levy mechanism.

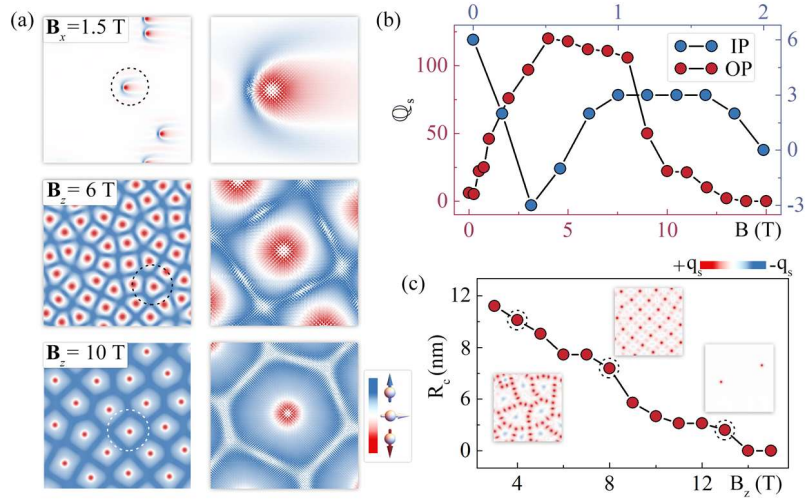

**Fig. 3.** (a) Spin textures of monolayer $Cr_2NSb$ under magnetic fields of $B_x$ = 1.5 T, $B_z$ = 6 T and $B_z$ =10 T. The right panels show the enlarged spin distributions. $S_i^z$ are denoted by colors and IP components of spins are indicated by arrows. (b) Evolutions of $\mathbb{Q}_s$ for monolayer $Cr_2NSb$ as functions of OP and IP external magnetic fields. (c) Evolution of radius of meron and skyrmion core $R_c$ as a function of OP magnetic field. Inserts in (c) present the corresponding $q_s$, and the color map specifies the sign of $q_s$.

In R-space topology, the ratio $|d/J|$ is a key criterion for the presence of topological magnetism. In detail, $|d/J| > 0.1$ indicates stable topological magnetism, while $|d/J| < 0.1$ suggests trivial FM phase [25, 26]. For monolayer $Cr_2NSb$, the ratios are estimated to be 0.05, 1.44 and 0.31, respectively, for Cr-Cr, Cr-Sb-Cr and Cr-N-Cr pairs. The substantially large ratios of NNN Cr atomic pairs can compensate for that of NN Cr atomic pair, enabling the possibility of stabilizing nontrivial topological magnetism in monolayer $Cr_2NSb$. To explore the spin textures of monolayer $Cr_2NSb$, the atomic spin model simulations are performed. As shown in **Fig. S4**, the meron-antimeron loops (MAL) emerge, serving as indication for multiple topological magnetism [33, 34].

To obtain multiple topological magnetism in monolayer $Cr_2NSb$, external magnetic fields are employed. By applying IP magnetic field $\mathbf{B}_x$ of 0.2 T, as shown in **Fig. S4**, meron and antimeron start to pair up, and bimerons emerge. Upon increasing $\mathbf{B}_x$ to 0.6 T, MAL is totally eliminated, and bimeron phase is realized in monolayer $Cr_2NSb$. The corresponding topological charge density $q_s$ is presented in **Fig. S5**, from which we can see that bimerons with both $\mathbb{Q} = 1$ and -1 coexist. This leads to small total topological charge $\mathbb{Q}_s$ [see **Fig. 3(b)**]. With increasing $\mathbf{B}_x$ to 1.5 T, as shown in the top panel of **Fig. 3(a)**, the bimeron phase consisting of bimerons with $\mathbb{Q} = 1$ is realized. The bimeron phase can be sustained with increasing $\mathbf{B}_x$ up to 2 T, and then transforms into trivial FM phase. Different from the case of $\mathbf{B}_x$, when applying OP magnetic field $\mathbf{B}_z$, merons and antimerons with $-q_s$ gradually expand to approximate the FM background [middle panel of **Fig. 3(a)**]. By increasing $\mathbf{B}_z$ larger than 8 T, as shown in the bottom panel of **Fig. 3(a)**, the skyrmion phase is obtained. With further increasing $\mathbf{B}_z$, the density of skyrmion decreases, and then reduces to zero under 14 T [**Fig. 3(b)**], implying a R-space topological transition to the trivial FM phase. Therefore, the topological magnetism of skyrmion and bimeron, as well as their switching, driven by magnetic field can be realized in monolayer $Cr_2NSb$.

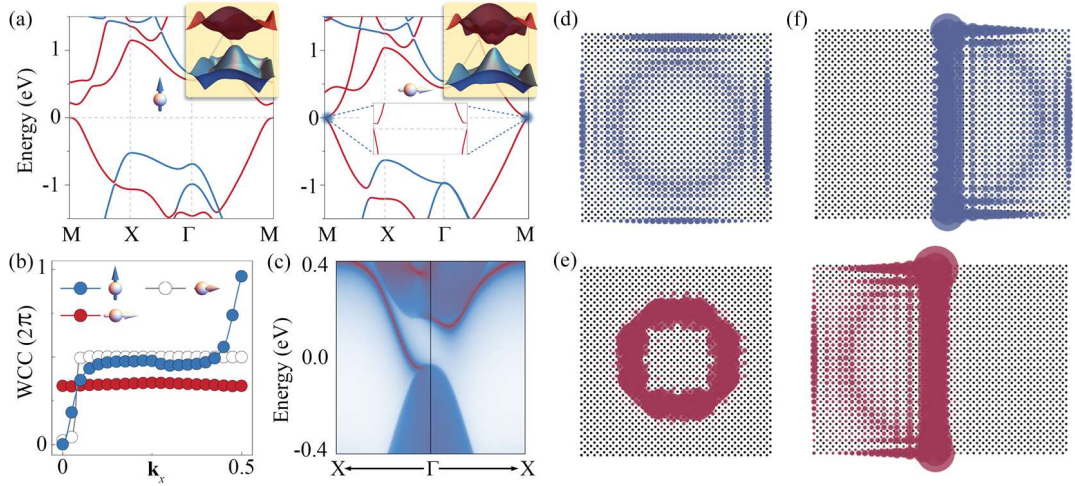

**Fig. 4.** (a) Band structures of OP-↑ and IP phases for monolayer $Cr_2NSb$ with SOC. Red and blue lines represent contributions from spin-up and spin-down bands, respectively. Inserts present the corresponding 3D plots of low-energy bands. Zoom-in band structures of IP phase are illustrated to capture the small gap. (b) Evolutions of WCCs for monolayer $Cr_2NSb$ along $\mathbf{k}_x$ with magnetization oriented along $\mathbf{z}$, $\mathbf{x}$ and $\mathbf{u}$, respectively. (c) Edge spectra of OP-↑ phase for monolayer $Cr_2NSb$. Distributions of CBSs of (d) zero-field CBS-M, (e) CBS-SK under $\mathbf{B}_z$=8 T and (f) CBS-BM under $\mathbf{B}_x$=1.5 T for monolayer $Cr_2NSb$.

After estimating the R-space topology, we then explore the K-space topology in monolayer $Cr_2NSb$. **Fig. S6** presents the band structure of monolayer $Cr_2NSb$ in the absence of SOC. Its spin-up bands exhibit a crossing point at the Fermi level, while spin-down bands show a large band gap. The crossing point at the M point is protected by the $C_{4v}$ point group of wave vector. For the OP phase, the inclusion of SOC deforms the symmetry protection and thus the crossing point, which results in a sizable band gap (~195 meV), as shown in **Fig. 4(a)**. **Fig. 4(b)** displays the evolution of Wannier charge centers (WCCs), with $k_y$ being treated as an adiabatic parameter. It can be seen that, with an arbitrary horizontal line, the WCC curve intersects odd times, which indicates the QAH state for the OP phase. This is further confirmed by the edge spectra of the OP phase presented in **Fig. 4(c)**, from which we can see a chiral gapless edge state appears within the bulk gap.

According to the previous works [36], the nontrivial topology strongly depends on spin orientations. The band structure of IP phase is depicted in **Fig. 4(a)**, from which we can see that a small gap (~5 meV) is generated. As the magnetization orientation rotates in the *xy* plane, monolayer $Cr_2NSb$ exhibits the joint $C_{2z}T$ symmetry (a two-fold rotation and time-reversal symmetry). $C_{2z}T$ symmetry enforces an odd Hall conductivity $\sigma_{xy}$, which guarantees $\mathbb{C}=0$. This is consistent with the WCCs of IP phase with spin oriented along **x** and **u** (**u** represents the bisector direction between **x** and **y**). These indicate that the IP phase is a trivial semiconductor. Therefore, monolayer $Cr_2NSb$ harbors the spin-orientation tunable QAH state.

At last, we investigate the interaction of topological magnetism and band topology, i.e., the multi-space topology, in monolayer $Cr_2NSb$. As illustrated by our model analysis, CBSs can describe the interactions of topological magnetism and QAH state, i.e., characterizing the multi-space topology. In real materials, the stability of CBSs is influenced by the size of topological magnetism. Specifically, when the size is considerably smaller than the penetration depth of QAH state, pronounced hybridization between edge and bulk states occurs, leading to the destruction of CBSs. The penetration depth of QAH state for monolayer $Cr_2NSb$ is estimated to be ~0.7 nm (see **Supplementary Note 5**). The radius of bimeron realized in monolayer $Cr_2NSb$ in the presence of $B_x$ is ~8 nm (**Fig. S4**). While for the radius of skyrmion in monolayer $Cr_2NSb$, with increasing $B_z$ from 8 to 13 T, it monstrously decreases from ~12 to ~6.4 nm, as shown in **Fig. 3(c)**. Obviously, the size of topological magnetism in monolayer $Cr_2NSb$ is much larger than the penetration depth of QAH state. Based on the detailed topological spin textures in monolayer $Cr_2NSb$, we further evaluate its CBSs according to **Eq. (3)**. The CBS-M under zero field, CBS-SK under $B_z$ = 8 T and CBS-BM under $B_x$=1.5 T are presented in **Figs. 4(d-f)**, respectively. Intriguingly, they are all strongly localized, and distributed significantly different. This, combined with the skyrmion-bimeron switching nature in monolayer $Cr_2NSb$ through magnetic field modulation, facilitates switchable multi-space topologies associated with different CBSs, thereby generating the coupling of multi-space topologies. And from **Figs. 4(d-f)**, it can be clearly seen that due to the significantly different distribution characters of these CBSs, this process is also accompanied with the space-shift of the CBSs.

**Conclusion**

In conclusion, we report the multi-space topology in 2D ferromagnetic lattice, arising from the interaction of topological magnetism and band topology. Using continuum theory and tight-binding model, we reveal that the interconnection between skyrmion/bimeron and QAH state can be manifested in distinctive localized CBSs. By modulating topological magnetism through magnetic field, the multi-space topologies associated with different CBSs can be switched, yielding the coupling of multi-space topologies. Such switching is accompanied with the space-shift of the CBSs. Using first-principles calculations and atomic spin model simulations, we further demonstrate the multi-space topologies and their strong coupling in monolayer $Cr_2NSb$.

**Acknowledgement**

We thank Runhan Li for helpful discussions on the QAH state of $Cr_2NSb$. This work is supported by the National Natural Science Foundation of China (Nos. 12274261 and 12074217), Taishan Young Scholar Program of Shandong Province and Shandong Provincial QingChuang Technology Support Plan (No. 2021KJ002).

**Data Availability**

The data that support the findings of this study are available within this article and its Supplementary Material.